\documentclass[aps,prl,twocolumn]{revtex4}
\usepackage{amsmath}
\usepackage{amssymb}
\usepackage{graphicx}

\input{tcilatex}

\begin{document}

\title{Surface acoustic wave induced magnetoresistance oscillations in a 2D
electron gas}
\author{John P. Robinson$^{1}$, Malcolm P. Kennett$^{2}$, Nigel R. Cooper$%
^{2}$, and Vladimir I. Fal'ko$^{1,3}$ }
\affiliation{$^{1}$ Physics Department, Lancaster University, LA1 4YB, UK}
\affiliation{$^{2}$ Cavendish Laboratory, University of Cambridge, Madingley Road,
Cambridge CB3 0HE, UK}
\affiliation{$^{3}$ Physics Department, Princeton University, Princeton, NJ 08544}
\date{\today}

\begin{abstract}
We study the geometrical commensurability oscillations imposed onto the
resistivity of 2D electrons in a perpendicular magnetic field by a
propagating surface acoustic wave (SAW). We show that, for $\omega <\omega
_{c}$, this effect is composed of an anisotropic dynamical classical
contribution increasing the resistivity and the non-equilibrium quantum
contribution isotropically decreasing resistivity, and we predict the
appearance of zero-resistance states associated with geometrical
commensurability at large SAW amplitude. We also describe how the
commensurability oscillations modulate the resonances in the SAW-induced
resistivity at multiples of the cyclotron frequency.
\end{abstract}

\maketitle

High-mobility two-dimensional electron gases (2DEGs) display interesting
effects under intense microwave irradiation \cite%
{Zudov,Mani,Durst,Andreev,Aleiner,Mirlin,KuKuMPL} or the influence of
surface acoustic waves in the microwave frequency range \cite{Willett,SAW}.
One class of effects is induced magneto-oscillations that originate from the
geometric commensurability between the cyclotron radius $R_{c}$ and the
period $2\pi /q$ of spatially periodic perturbations \cite{Pippard}, which
has been observed in statically modulated 2D systems \cite%
{Weiss,Beenakker,Gerhardts}. This effect was recently studied in the
attenuation and renormalization of surface acoustic wave (SAW) velocity due
to interactions with electrons \cite{Willett,Shilton} and in the drag effect 
\cite{Shilton}. For a spatially periodic propagating field of the SAW, the
commensurability effect can also be viewed as the resonant SAW interaction
with collective excitations of 2D electrons at finite wavenumbers \cite%
{Willett,Halperin}, enabling one to excite modes otherwise forbidden by
Kohn's theorem \cite{Kohn}. 

In this Letter, we study the non-linear dynamical effect in which SAWs
induce changes in the magneto-resistivity of a high quality electron gas in
the regime of classically strong magnetic fields, $\omega _{c}\tau \gg 1$
and high temperatures $k_BT\gg \hbar \omega _{c}$. We show that the
resistivity changes reflect both frequency and geometrical resonances in the
surface acoustic wave attenuation and are formed from two competing
contributions.

The first contribution originates from the SAW-induced guiding center drift
of the cyclotron orbits -- a purely classical effect. For a SAW with
frequency $\omega $, and wavenumber $q$, propagating in the $x$ direction
with speed $s=\omega /q$, there is an anisotropic increase in the
resistivity $\rho _{xx}$ (at high fields $\omega _{c}\tau \gg 1,$ this is
equivalent to an increase of conductivity in the transverse direction $%
\sigma _{yy}$), which oscillates as a function of inverse magnetic field
when the Fermi velocity $\mathrm{v}_{F}\gg s$. We show that at $\omega
\gtrsim \omega _{c}$ the resistivity change displays resonances at multiples
of the cyclotron frequency $\omega \approx N\omega _{c}$. 

The second contribution arises from the modulation of the electron density
of states (DOS), $\tilde{\gamma}(\epsilon )=\left[ 1-\Gamma \cos \left( 2\pi
\epsilon /\hbar \omega _{c}\right) \right] \gamma $ (where $\gamma =m/\pi
\hbar ^{2}$), and consequently, from the energy dependence of the
non-equilibrium population of excited electron states caused by Landau level
quantization. We follow the idea proposed in Ref.~\cite{Aleiner} to explain
the formation of zero-resistance states \cite{Zudov,Mani,Andreev} under
microwave irradiation with $\omega \gtrsim \omega _{c}$. We show that in the
frequency range $\tau ^{-1}\lesssim \omega \lesssim \omega _{c}$ the quantum
contribution suppresses resistivity both in $\rho _{xx}$ and $\rho _{yy}$
and persists up to temperatures $k_{B}T\gg \hbar \omega _{c}$ and filling
factors $\nu \gg 1$ where no Shubnikov-de Haas oscillations would be seen in
the linear-response conductivity.

When $\tau ^{-1}\lesssim \omega \ll \omega _{c}$, the commensurability
oscillations take the form 
\begin{eqnarray}
\frac{\delta \rho _{xx}}{\rho _{xx}} &\approx &\frac{\delta \sigma _{yy}}{%
\sigma _{xx}}=2J_{0}^{2}\left( qR_{c}\right) \mathcal{E}^{2}\left[ \frac{%
\mathrm{v}_{F}^{2}}{s^{2}}-\frac{\tau _{in}}{\tau }\left( 2\pi \Gamma \nu
\right) ^{2}\right] ,  \notag \\
\frac{\delta \rho _{yy}}{\rho _{xx}} &\approx &\frac{\delta \sigma _{xx}}{%
\sigma _{xx}}=-2J_{0}^{2}\left( qR_{c}\right) \mathcal{E}^{2}\times \frac{%
\tau _{in}}{\tau }\left( 2\pi \Gamma \nu \right) ^{2},  \label{FullResult}
\end{eqnarray}%
with $\mathcal{E}=ea_{\mathrm{scr}}E_{\omega q}^{\mathrm{SAW}}/\epsilon _{F} 
$, $E_{\omega q}^{\mathrm{SAW}}$ the SAW longitudinal electric field, $a_{%
\mathrm{scr}}=\chi /2\pi e^{2}\gamma$ the 2D screening radius, $\epsilon
_{F} $ the Fermi energy, $\tau $ and $\tau _{in}$ the momentum and inelastic
relaxation times \cite{footnote}, and $\nu =2\epsilon _{F}/\hbar \omega _{c}$
the filling factor. The geometrical oscillations in Eq.~(\ref{FullResult})
are described by the Bessel function $J_{0}$. For different sample
parameters and measurement conditions, the observed oscillatory change in
the resistivity can be dominant in the SAW propagation direction [for $\eta
\equiv\left( 2\pi \Gamma \nu s/\mathrm{v}_{F}\right) \sqrt{\tau _{in}/\tau }%
<1$], isotropic [$\eta \gg 1$], or only in the component perpendicular to
the SAW wavevector [if $\eta \approx 1$].

\begin{figure}[h]
\center{\includegraphics[width=8.3cm]{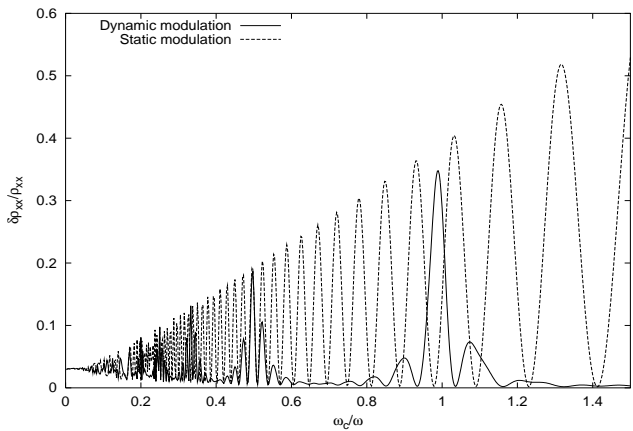}}
\caption{Magnetoresistance for both static ($\protect\omega \protect\tau =0$%
) and dynamic ($\protect\omega \protect\tau =20$) modulations of the 2DEG,
for $ql=600$, and ${\mathcal{E}}=0.01$ ($l=v_{F}\protect\tau $ is the mean
free path). Both the cyclotron harmonics and the geometrical
commensurability oscillations are visible for these parameters.}
\label{fig:magr}
\end{figure}

We now present our detailed analysis, starting with the \textit{dynamical
classical contribution}. At high magnetic fields $\omega _{c}\tau \gg 1$,
the resistivity change $\delta ^{c}\rho _{xx}$ can be tracked back to the
SAW induced drift $Y(t)$ (along the $y$-axis) of the guiding centre of an
electron cyclotron orbit \cite{Beenakker} and the resulting enhancement of
the transverse ($y$-) component of the electron diffusion coefficient, 
\begin{equation*}
\dfrac{\delta ^{c}\rho _{xx}}{\rho _{xx}}=\dfrac{\delta D_{yy}}{D_{yy}}\sim 
\dfrac{\langle Y^{2}\rangle /\tau }{R_{c}^{2}/\tau }=\left\langle \left( 
\dfrac{Y}{R_{c}}\right) ^{2}\right\rangle .
\end{equation*}%
The drift is caused by an electric field $E_{\omega q}\cos (qx-\omega t+\phi
)\hat{\mathbf{x}}$, with $x(t)=R_{c}\sin \left( \omega _{c}t-\psi \right) $.
Between two impurity scattering events the guiding centre is displaced by 
\begin{equation*}
Y(t)\sim \frac{eE_{\omega q}}{m\omega _{c}}\int_{0}^{t}d\tilde{t}\cos \left[
qR_{c}\sin (\omega _{c}\tilde{t}-\psi )-(\omega \tilde{t}-\phi )\right] .
\end{equation*}%
A particular guiding centre displacement $Y(t)$ depends on the initial phase 
$\psi $ of the electron revolution along the cyclotron orbit and the phase $%
\phi $ of the SAW field which change randomly each time electron scatters
form impurities, if $R_{c}>2\pi /q,$ thus leading to a random change in the
value and direction of $Y$. The mean square value of such a displacement
(averaged over $\psi $, $\phi $ and the drift time between two scattering
events), 
\begin{equation}
\langle Y^{2}\rangle =\int_{0}^{\infty }\dfrac{dt}{\tau }e^{-t/\tau
}\int_{0}^{2\pi }\int_{0}^{2\pi }\dfrac{d\psi }{2\pi }\dfrac{d\phi }{2\pi }%
Y^{2}(t),
\end{equation}%
can be used to obtain the frequency and wave number dependence of the
effect, 
\begin{equation}
\frac{\delta ^{c}\rho _{xx}}{\rho _{xx}}=2\left( ql\mathcal{E}\right)
^{2}\sum_{N=-\infty }^{\infty }\frac{J_{N}^{2}(qR_{c})}{1+\left( \omega
-N\omega _{c}\right) ^{2}\tau ^{2}},  \label{Fullresult2}
\end{equation}%
where $\mathcal{E}=ea_{\mathrm{scr}}E_{\omega q}^{\mathrm{SAW}}/\epsilon
_{F} $ and $l=\mathrm{v}_{F}\tau $. Equation~(\ref{Fullresult2}) includes
the Thomas-Fermi screening of the SAW field by 2D electrons, $E_{\omega
q}=qa_{\mathrm{scr}}E_{\mathbf{\omega }q}^{\mathrm{SAW}}$. A typical change
in the magnetoresistance $\delta ^{c}\rho _{xx}$ for the regime $qR_{c}\gg {%
\omega }/{\omega _{c}}\gtrsim 1$ (possible at normal electron densities for
which $\mathrm{v}_{F}\gg s$) is illustrated in Fig.~\ref{fig:magr} and
compared with the Weiss oscillations in a static potential.

There are a sequence of resonances at integer multiples of the cyclotron
frequency, $\omega \approx N\omega _{c}$. The oscillations for even
harmonics are in phase with the Weiss oscillations of the static potential,
whilst those of odd harmonics are $\pi$ out of phase. This is because the
main contributions to the drift occur when the electrons are moving parallel
(or anti-parallel) to equipotential lines. For odd harmonics the phase of
the potential at the half-orbit point is opposite to that for a static
potential, and hence cancellation and reinforcement effects that lead to
minima and maxima in the resistance are interchanged. In the regime $\omega
_{c}\tau \gg 1$, the resonances are very narrow and appear to display a
random sequence of heights, reflecting the dependence on the geometric
resonance conditions.

In the intermediate frequency domain, $\tau ^{-1}\ll \omega \ll \omega _{c}$%
, a natural regime for GaAs structures with densities $n_{e}\gtrsim 10^{10}%
\mathrm{cm}^{-2}$ at sufficiently high magnetic fields, the classical
oscillations take the form 
\begin{equation}
\frac{\delta ^{c}\rho _{xx}}{\rho _{xx}}\approx \frac{2\mathrm{v}_{F}^{2}}{%
s^{2}}\mathcal{E}^{2}J_{0}^{2}(qR_{c}).  \label{saturation}
\end{equation}%
Due to the competition between electron screening effects ($\propto q^{2}$),
the dynamical suppression of commensurability by the SAW motion ($\propto
\omega ^{-2}$), and the relation $\omega /q=s$, the form of these
oscillations is independent of the absolute value of the SAW frequency,
provided that conditions $\tau ^{-1}\lesssim\omega $ and $v_F \gg s$ are
satisfied.

The dynamical mechanism just described dominates in a classical electron
gas. No redistribution of electron kinetic energy (due to SAW absorption)
will additionally change the magnetoresistance until the 2DEG is heated to a
temperature $k_{B}T_{e}\gtrsim \hbar \omega _{c}k_{F}/q$ where geometrical
oscillations become smeared. The essential assumption leading to this
statement is that electron single-particle parameters (velocity and $\tau
^{-1}$) vary slowly with energy at the scales comparable to the Fermi energy
and can thus be approximated by constants.

To demonstrate this point, we analyze the classical kinetic equation for a
2DEG at temperature $k_{B}T\lesssim\hbar \omega _{c}k_{F}/q$ irradiated by
SAWs. We solve the kinetic equation for the electron distribution function, 
\begin{equation}
f(t,x,\varphi ,\epsilon )=f_{T}+\sum\limits_{\omega q}e^{-i\omega
t+iqx}\sum_{m}f_{\omega q}^{m}(\epsilon )e^{im\varphi },
\end{equation}%
using the method of successive approximations. Here, $f_{T}(\epsilon )$ is
the homogeneous equilibrium Fermi function, and the angle $\varphi $ and
kinetic energy $\epsilon $ parametrize the electron state in momentum space.
Each component $f^{m}$ describes the $m$-th angular harmonic of the time-
and space-dependent non-equilibrium distribution. To describe local values
of the electron current and the accumulated charge density, we use the
energy-integrated functions, $g_{\omega q}^{m}=\int_{0}^{\infty }d\epsilon
f_{\omega q}^{m}$. The relaxation of the local non-equilibrium distribution
towards a Fermi function characterized by the value of local Fermi energy, $%
\epsilon _{F}(t,x)$ (determined by the local electron density $n(t,x)\propto
g^{0}(t,x)=\int_{0}^{\infty }d\epsilon f^{0}(t,x)$, where $f^{0}(t,x)=\int 
\dfrac{d\varphi }{2\pi }f(t,x)$) is described in the relaxation-time
approximation by 
\begin{equation}
\mathcal{\hat{L}}f=-\dfrac{f-f^{0}}{\tau }-\dfrac{f^{0}-f_{T}(\epsilon
-\epsilon _{F}(t,x))}{\tau _{in}},  \label{KinEq}
\end{equation}%
\begin{equation*}
\mathcal{\hat{L}}=\partial _{t}+\mathrm{v}\cos \varphi \partial _{x}+\left[
\omega _{c}-\frac{eE}{p}\sin \varphi \right] \partial _{\varphi }+e\mathrm{v}%
E\cos \varphi \partial _{\epsilon },
\end{equation*}%
where we distinguish between the elastic scattering rate $\tau ^{-1}$ and
energy relaxation rate $\tau _{in}^{-1}$. The electric field $E$ in Eq.~(\ref%
{KinEq}) is the combination of a homogeneous DC field $\mathbf{E}_{00}$ and
the screened electric field of the SAW, $\mathbf{E}(t,x)=\sum_{\omega
q}E_{\omega q}e^{iqx-i\omega t}\hat{\mathbf{x}}$, found from the unscreened
SAW field via $E_{\omega q}=E_{\omega q}^{\mathrm{SAW}}/\kappa (q,\omega )$,
where $\kappa (q,\omega )$ is the dielectric function of the whole 2D
structure.

The dynamical perturbation of the distribution function can be found from
time/space Fourier harmonics of Eq. (\ref{KinEq}) at the frequency/wave
number of the SAW, 
\begin{equation}
\lbrack \partial _{\varphi }+\frac{1}{\omega _{c}\tau }-i\frac{\omega }{%
\omega _{c}}+iqR_{c}\cos \varphi ]f_{\omega q}=\Psi (\varphi ),
\label{KinHarm}
\end{equation}
\begin{eqnarray*}
\Psi &=&-\frac{g_{\omega q}^{0}}{\omega _{c}\tau _{in}}(\partial _{\epsilon
}f_{T})+\dfrac{\tau ^{-1}-\tau _{in}^{-1}}{\omega _{c}}f_{\omega q}^{0} \\
&&-\frac{eE_{\omega q}}{\omega _{c}}[\mathrm{v}\cos \varphi \partial
_{\epsilon }-\frac{\sin \varphi }{p}\partial _{\varphi }](f_{00}+f_{T}),
\end{eqnarray*}%
where we include the unknown perturbation of the time/space averaged
function, $f_{00}$, related to the DC current to lowest order in $E_{\omega
q}$. Equation (\ref{KinHarm}) can be formally solved using the Green
function $G(\varphi ,\tilde{\varphi})$ 
\begin{eqnarray}
f_{\omega q}(\varphi ) &=&\int_{-\infty }^{\varphi }G(\varphi ,\tilde{\varphi%
})\Psi (\tilde{\varphi})d\tilde{\varphi},  \label{GF1} \\
G(\varphi ,\tilde{\varphi}) &=&e^{[\frac{1}{\omega _{c}\tau }-\frac{i\omega 
}{\omega _{c}}](\tilde{\varphi}-\varphi )+iqR_{c}[\sin \tilde{\varphi}-\sin
\varphi ]}\,,  \label{GF2}
\end{eqnarray}%
which allows for an infinite range of variation of $\varphi $ whilst
guaranteeing periodicity of the solution $f_{\omega q}(\varphi )$.

Thomas-Fermi screening of the SAW field is produced by the density
modulation $n_{\omega q}=\gamma g_{\omega q}^{0}$ via the induced field $%
E_{\omega q}^{\mathrm{ind}}=-\frac{i2\pi eq}{\chi |q|}\gamma g_{\omega q}^{0}
$ , so that $E_{\omega q}=E_{\omega q}^{\mathrm{SAW}}+E_{\omega q}^{\mathrm{%
ind}}$. In the analysis of screening, the DC part of the electric field can
be ignored ($f_{00}=0$), and self-consistency yields 
\begin{eqnarray}
g_{\omega q}^{0} &=&\frac{eE_{\omega q}}{iq}\frac{1-(1-i\omega \tau )K}{1-K},
\label{g-naught} \\
K &=&\int_{0}^{2\pi }\dfrac{d\varphi }{2\pi }\int_{-\infty }^{\varphi }%
\dfrac{d\tilde{\varphi}}{\omega _{c}\tau }G(\varphi ,\tilde{\varphi})  \notag
\\
&=&\sum_{N=-\infty }^{\infty }\dfrac{J_{N}^{2}(qR_{c})}{1+i\tau (N\omega
_{c}-\omega )}\quad   \label{K-1}
\end{eqnarray}%
where $K(\omega ,q)$ obeys the relations $K(\omega ,q)=K^{\ast }(-\omega
,q)=K(\omega ,-q)$ and is obtained from Eqs.~(\ref{GF1},\ref{GF2}) using the
identity $e^{iz\sin \varphi }=\sum_{N=-\infty }^{\infty
}J_{N}(z)e^{iN\varphi }$, and 
\begin{equation}
\kappa =1+\frac{1}{a_{\mathrm{scr}}|q|}\frac{1-\left( 1-i\omega \tau \right)
K}{1-K}.  \label{kappa}
\end{equation}

To find the steady state current, we analyze the time/space average of the
kinetic equation in Eq. (\ref{KinEq}) and take into account the dynamical
perturbation $f_{\omega q}$, 
\begin{eqnarray}
&&[\partial _{\varphi }+\frac{1}{\omega _{c}\tau }]f_{00}-\dfrac{\tau
^{-1}-\tau _{in}^{-1}}{\omega _{c}}f_{00}^{0}+\frac{e\mathbf{v\cdot E}_{00}}{%
\omega _{c}}\partial _{\epsilon }f_{T}  \notag \\
&=&-\sum_{\omega q}\frac{eE_{-\omega -q}}{\omega _{c}}[\mathrm{v}\cos
\varphi \partial _{\epsilon }-\frac{\sin \varphi }{p}\partial _{\varphi
}]f_{\omega q}.  \label{zeroharm}
\end{eqnarray}%
We substitute the solution Eq.~(\ref{GF1}) into Eq.~(\ref{zeroharm}),
keeping track of the effect of the perturbation of the time/space averaged
function $f_{00}$ on $f_{\omega q}$. We thus include SAW-induced non-linear
effects. We multiply Eq.~(\ref{zeroharm}) by $(2\omega _{c}/\mathrm{v}%
)e^{-i\varphi }$, integrate with respect to $\epsilon $ and $\varphi $, then
use the relation between the $x$ and $y$ components of the DC current, $%
j_{x}-ij_{y}=e\gamma \mathrm{v}_{F}g_{00}^{1}$ and the harmonic $g_{00}^{1}$
(note that electrical neutrality requires $g_{00}^{0}=0$), to get%
\begin{equation*}
\begin{array}{c}
\frac{2\omega _{c}}{e\mathrm{v}_{F}^{2}}\int \frac{d\varphi }{2\pi }%
e^{-i\varphi }\int d\epsilon \left\{ \lbrack \partial _{\varphi }+\frac{1}{%
\omega _{c}\tau }]\mathrm{v}f_{00}\right. \\ 
\left. +\sum_{\omega q}\frac{\mathrm{v}eE_{-\omega -q}}{\omega _{c}}[\mathrm{%
v}\cos \varphi \partial _{\epsilon }-\frac{\sin \varphi }{p}\partial
_{\varphi }]f_{\omega q}\right\} \\ 
=\frac{j_{x}}{\mathrm{v}_{F}^{2}\tau e^{2}\gamma /2}\times \frac{1}{2}%
\sum_{\omega q}\left\vert \frac{leE_{\omega q}}{\epsilon _{F}}\right\vert
^{2}\frac{K}{1-K}+ \\ 
+\left( \frac{i\omega _{c}\tau +1}{\mathrm{v}_{F}^{2}\tau e^{2}\gamma /2}%
\right) [j_{x}-ij_{y}]=E_{00}^{x}-iE_{00}^{y},%
\end{array}%
\end{equation*}
which can be used to determine the SAW induced change of the resistivity
tensor, $\delta ^{c}\hat{\rho}$. The relation between the electric field and
current is $\mathbf{E}=\hat{\rho}\,\mathbf{j}+\delta ^{c}\hat{\rho}\,\mathbf{%
j}$, where $\hat{\rho}$ is the Drude resistivity tensor, thus we find the
resistivity corrections 
\begin{eqnarray}
\frac{\delta ^{c}\rho _{xx}}{\rho _{xx}} &=&\frac{1}{2}\sum_{\omega
q}\left\vert \frac{elE_{\omega q}}{\epsilon _{F}}\right\vert ^{2}\mathrm{Re}%
\left\{ \frac{K}{1-K}\right\} ,  \label{WithK} \\
\delta ^{c}\rho _{yy} &=&0,  \notag \\
\dfrac{\delta ^{c}\rho _{yx}}{\rho _{xx}} &=&\dfrac{\delta ^{c}\rho _{xy}}{%
\rho _{xx}}=-\frac{1}{2}\left\vert \frac{leE_{\omega q}}{\epsilon _{F}}%
\right\vert ^{2}\mathrm{Im}\sum_{\omega q}\frac{K}{1-K}=0,  \notag
\end{eqnarray}%
which, with the use of $E_{\omega q}=E_{\omega q}^{\mathrm{SAW}}/\kappa
(q,\omega )$ and Eq. (\ref{K-1}), yields the result in Eq. (\ref{Fullresult2}%
).

The magnetic field dependence of the resistivity change reflects the form of
the SAW attenuation by the 2D electrons determined by the real part of the
longitudinal dynamical conductivity $\sigma _{\omega q}$, 
\begin{equation}
\mathrm{Re}\,\sigma _{\omega q}=\gamma s^{2}\tau e^{2}\,\mathrm{Re}\left\{ 
\frac{K}{1-K}\right\} \mathrm{.}  \label{SigmaOmegaQ}
\end{equation}%
The finite wavenumber $q>8a_{\mathrm{scr}}/R_{c}^{2}$ of the SAW allows the
system to bypass Kohn's theorem and generates absorption resonances at \cite%
{deltaN} $\omega =N\omega _{c}+\Delta _{N}(q)$, 
\begin{equation}
\left\langle \mathbf{E\cdot j}\right\rangle =\frac{\gamma \left\vert eqa_{%
\mathrm{scr}}E_{\omega q}^{\mathrm{SAW}}\right\vert ^{2}s^{2}\tau }{1+\tau
^{2}\left( \omega -N\omega _{c}-\Delta _{N}\right) ^{2}}J_{N}^{2}(qR_{c}). 
\notag
\end{equation}

Absorption of SAWs by the 2D electrons changes their steady state
distribution over energy, though for energy-independent characteristics this
does not lead to additional changes in magnetoresistance beyond those
described in Eq.~(\ref{saturation}). However, Landau level quantization,
which is unavoidable in a phase-coherent electron gas when $\omega _{c}\tau
\gg 1$, gives rise to the \textit{quantum contributions to the geometrical
commensurability oscillations} which persist up to high temperatures. The
oscillatory energy-dependence of the electron DOS, $\tilde{\gamma}(\epsilon
)=\left[ 1-\Gamma \cos \left( 2\pi \epsilon /\hbar \omega _{c}\right) \right]
\gamma $, imposes oscillations on the electron elastic scattering rate, $%
\tau ^{-1}(\epsilon )=\tau ^{-1}\tilde{\gamma}/\gamma $, and the
contribution to the observable conductivity \cite{Mirlin}, 
\begin{equation*}
S(\epsilon )\approx \sigma _{xx}\tilde{\gamma}^{2}/\gamma ^{2} , \; \quad
\sigma =\int d\epsilon \, S(\epsilon )[-\partial _{\epsilon }f_{00}^{0}].
\end{equation*}%
At low temperatures, $k_{B}T\lesssim \hbar \omega _{c}$, the DOS
oscillations lead to Shubnikov-de Haas oscillations in conductivity. At high
temperatures $k_{B}T\gg \hbar \omega _{c}$, thermal broadening smears out
oscillations, but the quantum contribution can remain in non-linear effects
after energy averaging.

It was shown in Ref.~\cite{Aleiner} that the electron energy distribution
acquires an oscillatory part via the availability of final states for energy
absorption processes. When the electron gas is excited by SAWs, carriers are
redistributed between energy intervals with varying transport efficiencies,
changing the overall resistivity of the 2DEG in proportion to the SAW
attenuation. We extend the study of the balance equation performed in Refs.~%
\cite{Aleiner,Mirlin} for the non-equilibrium electron distribution (at $%
\omega \gtrsim \omega _{c}$) to low-frequencies $\omega \ll \omega _{c}$ 
\cite{footnote2}, where it has the form \ 
\begin{equation*}
\sum\limits_{\pm \omega q}\left\vert E_{\omega q}\right\vert ^{2}\frac{%
\sigma _{\omega q}}{\gamma \tilde{\gamma}}\partial _{\epsilon }\left[ \frac{%
\tilde{\gamma}^{2}}{\gamma ^{2}}\partial _{\epsilon }\left(
f_{00}^{0}+f_{T}\right) \right] =\frac{f_{00}^{0}}{\tau _{in}}.
\end{equation*}%
The solution of this equation splits into smooth and oscillatory
energy-dependent parts, $f_{00}^{0}=\left\langle f_{00}^{0}\right\rangle
_{\epsilon }+\delta f^{0}(\epsilon )$. Here, $\left\langle \cdot \cdot \cdot
\right\rangle _{\epsilon }$ stands for averaging over a small energy scale $%
\sim \hbar \omega _{c}$. To lowest order in $\left\vert E_{\omega
q}\right\vert ^{2}$, the oscillatory part is 
\begin{equation*}
\delta f^{0}=\tau _{in}\sum\limits_{\pm \omega q}\left\vert E_{\omega
q}\right\vert ^{2}\frac{\sigma _{\omega q}}{\gamma \tilde{\gamma}}\partial
_{\epsilon }\left[ \frac{\tilde{\gamma}^{2}}{\gamma ^{2}}\partial _{\epsilon
}f_{T}\right] ,
\end{equation*}%
where $\sigma _{\omega q}$ is the longitudinal component of classical
non-local conductivity given by Eq.~(\ref{SigmaOmegaQ}) and $\tau
_{in}^{-1}\ll \tau ^{-1}$. We thus estimate the SAW-induced change in the
conductivity, $\delta ^{q}\sigma _{xx}=\int d\epsilon \,\delta f^{0}\left(
\partial _{\epsilon }S\right) $, as 
\begin{equation*}
\delta ^{q}\sigma _{xx}=\tau _{in}\sigma _{xx}\sum\limits_{\pm \omega
q}\left\vert E_{\omega q}\right\vert ^{2}\frac{\sigma _{\omega q}}{\gamma
^{5}}\int d\epsilon \partial _{\epsilon }\left[ \tilde{\gamma}^{2}\partial
_{\epsilon }f_{T}\right] \left( \partial _{\epsilon }\tilde{\gamma}%
^{2}\right) .
\end{equation*}

In the latter expression, the oscillatory energy dependence of the DOS,
which is enhanced by differentiation, $\partial _{\epsilon }\tilde{\gamma}%
=\gamma \Gamma \left( 2\pi /\hbar \omega _{c}\right) \sin \left( 2\pi
\epsilon /\hbar \omega _{c}\right) $, appears under the square, so that its
averaged broad thermal smearing gives $\left\langle \left( \partial
_{\epsilon }\tilde{\gamma}\right) ^{2}/\gamma ^{2}\right\rangle _{\epsilon }=%
\frac{1}{2}\left( 2\pi \Gamma /\hbar \omega _{c}\right) ^{2}$. This results
in a non-vanishing addition to both diagonal components of the conductivity
(and, therefore, also of the resistivity) even when $k_{B}T\gg \hbar \omega
_{c}$. This generates isotropic magneto-oscillations 
\begin{equation}
\dfrac{\delta ^{q}\rho _{\alpha \alpha }}{\rho _{xx}}=\dfrac{\delta
^{q}\sigma _{\alpha \alpha }}{\sigma _{xx}}=-\frac{2\tau _{in}}{\tau }%
\left\vert \frac{4\pi \Gamma \epsilon _{F}}{\hbar \omega _{c}}\right\vert
^{2}\mathcal{E}^{2}J_{0}^{2}\left( qR_{c}\right) ,
\end{equation}%
in addition to the anisotropic classical commensurability effect and,
together, they yield the result in Eq. (\ref{FullResult}).

In conclusion, we have demonstrated a new class of magnetoresistance
oscillations caused in a 2DEG by SAW. We have shown that the effect consists
of contributions with competing signs: (i) a classical geometric
commensurability effect analagous to that found in static systems with
positive sign, and (ii) a quantum correction, with negative sign. The latter
result suggests that SAW propagation through a high mobility electron gas
may generate a sequence of zero-resistance states (ZRS) linked to the maxima
of $J_{0}^{2}\left( qR_{c}\right) $ for strong enough SAW fields. Whilst
this prediction concerns the low-frequency domain $\omega \lesssim \omega
_{c}$, such ZRS would be formed via the same mechanism \cite{Andreev,Aleiner}
as the microwave-induced ZRS at $\omega \gtrsim \omega _{c}$. A large enough
SAW-induced change $\left\vert \delta \sigma _{xx}\right\vert >\sigma _{xx}$
resulting in negative local conductivity would require formation of electric
field/Hall current domains. Since the anisotropy in Eq.~(\ref{FullResult})
suggests that such conditions can be achieved the easiest in the
conductivity component along the SAW wavevector, we expect that domains
would form with current flowing perpendicular to the direction of SAW
propagation, and their stability would depend on the sample geometry \cite%
{footnote3}.

We thank I. Aleiner and A. Mirlin for stimulating discussions. This work was
funded by EPSRC grants GR/R99027 and GR/R17140 and progressed during the
workshop in the Max-Planck-Institut PKS in Dresden.


\begin{references}
\bibitem{footnote} We choose short-range disorder for computational convenience, and
 expect that long-range disorder will introduce quantitative changes to our results.
\bibitem{footnote2} We can only treat the quantum correction using the classical $\sigma_{\omega q}$
when $\omega \ll \omega_c$.
\bibitem{footnote3} For a SAW with the wavevector directed across the axis of a Hall bar, 
current domains can be stabilized by ending in ohmic contacts. 
For a wave propagating (or standing) along the Hall bar, 
current domains would have to orient across the bar direction 
and terminate at the sample edges (destabilizing them), 
leading to a finite resistance. 
Finally, the anisotropy would not support a zero-conductance 
regime in a Corbino geometry.
\end{references}
\end{document}